# Scanning Thermo-ionic Microscopy: Probing Nanoscale Electrochemistry via Thermal Stress-induced Oscillation

Ehsan Nasr Esfahani [1,2], Ahmad Eshghinejad [1], Yun Ou [2], Jinjin Zhao [3], Stuart Adler [4], and Jiangyu Li [1,2,*]

1. Department of Mechanical Engineering, University of Washington, Seattle, WA 98195, USA
2. Shenzhen Key Laboratory of Nanobiomechanics, Shenzhen Institutes of Advanced Technology, Chinese Academy of Sciences, Shenzhen 518055, Guangdong, China
3. School of Materials Science and Engineering, Shijiazhuang Tiedao University, Shijiazhuang, 050043, China
4. Department of Chemical Engineering, University of Washington, Seattle, WA 98195, USA

*jjli@uw.edu

## Abstract

A universal challenge facing the development of electrochemical materials is our lack of understanding of physical and chemical processes at local length scales in 10-100 nm regime, and acquiring this understanding requires a new generation of imaging techniques. In this article, we introduce the scanning thermo-ionic microscopy (STIM) for probing local electrochemistry at the nanoscale, using for imaging the Vegard strain induced via thermal stress excitations. Since ionic oscillation is driven by the stress instead of voltage, the responses are insensitive to the electromechanical, electrostatic, and capacitive effects, and they are immune to global current perturbation, making *in-operando* testing possible.

## Introduction

The worldwide energy economy is shifting focus from extraction and consumption of fossil fuels toward a diverse, multidirectional, and asynchronous network of energy sources and demands, and there will be a critical need for electrochemical materials that efficiently interconvert and/or store electrical and chemical energy. These include more robust electrode materials for large and small scale battery systems [1-3], fuel cells or flow batteries [4-6], electrocatalysts for efficient electrosynthesis of liquid transportation and storage fuels [7], and photoelectrochemical materials that can directly convert solar energy to fuels [8]. At present, a universal challenge facing the development of electrochemical materials is our lack of understanding of physical and chemical processes at local length scales in the 10-100 nm regime. A growing body of research shows that the composition, structure, and properties of electrochemical materials near active interfaces often deviate substantially and inhomogeneously from those of the bulk, and the electrochemistry at this length scale is still poorly understood [5, 9, 10]. Acquiring this understanding requires a new generation of imaging techniques that can resolve local chemistry and fast dynamics in electrochemical materials at the time along length scales relevant to strongly coupled reaction and transport phenomena. This offers a new opportunity for microscopy development.

Conventional electrochemical characterization techniques are very difficult to scale down as they are mostly based on the measurement of current, requiring detection of small currents on the order of pA at the nanoscale [11]. Custom-made ion-conducting electrodes have been developed for scanning electrochemical microscopy [12], though its spatial resolution is usually no better than micrometers [13, 14], and the fabrication process is rather complicated. Electrostatic force microscopy and Kelvin probe force microscopy have been applied to study local electrochemical processes [11, 15], yet the spatial resolution is limited there because of long-range electrostatic interactions. In addition the data are often not unambiguous and thus difficult to interpret. In the last several years, researchers have realized that Vegard strain [16, 17] can provide an alternative imaging mechanism with high sensitivity and spatial resolution for nanoscale electrochemical characterization. The initial attempt focused on atomic force microscopy (AFM) topographic mapping of lithium ion electrodes during charging and discharging, however the volume evolution that is induced by the change in lithium ion concentration [18] reflects the accumulation of Vegard strain over both space and time. Later on, electrochemical strain microscopy (ESM) was proposed, focusing on local and instantaneous fluctuations in ionic species induced by an AC voltage applied through a conductive scanning probe tip [19-22]. While these techniques have provided considerable insight into local electrochemistry, the measured strain is electromechanical in nature, and thus it is often difficult to distinguish Vegard strain from other electromechanical mechanisms such as piezoelectric effect, electrostatic interactions, and capacitive forces [23, 24]. Furthermore, it is highly challenging to integrate local ESM with global electrochemical measurement *in-operando* because the charged scanning probe is often affected by global voltage perturbations.

In order to overcome some of the difficulties associated with ESM, we have recently developed the scanning thermo-ionic microscopy (STIM) technique for probing local electrochemistry at the nanoscale [25, 26]. This new technique also utilizes Vegard strain for imaging, though the strain is induced via thermal stress excitations instead of an AC electric potential, thus eliminating the contributions from other electromechanical strains as well as electrical interferences from global voltage perturbations. This makes possible *in-operando* electrochemical testing at the nanoscale. We have implemented this technique using both resistive heating through a microfabricated AFM thermal probe and photo-thermal heating through a 405 nm laser, and we have applied it to a variety of electrochemical materials for imaging and spectroscopy studies. This article describes the principle of STIM and its implementation, its applications to ceria and perovskite solar materials, and some challenges and future developments.

**Materials and Methods**

**Concept of the method.** The concept of STIM is built on three observations. First, when the ionic concentration oscillates in a solid, the associated volume will fluctuate as well because of the so-called Vegard strain [16, 17], defined broadly as a lattice volume change associated with a concentration change in the ionic species. This in turn results in a mechanical vibration that can be measured locally via a scanning probe, shown schematically Figure 1. Such dynamic-strain-based detection has been applied in a variety of scanning probe microscopy (SPM) techniques, including piezoresponse force microscopy (PFM) [27-30], electrochemical strain microscopy (ESM) [19-21], and piezomagnetic force microscopy (PmFM) [31, 32]. Secondly, the fluctuation of ionic concentration can be driven by an oscillation in its electrochemical potential, which can be induced by gradient in ionic concentration, electric potential, or mechanical stress. Finally, the characteristics of ionic oscillation and electrochemical strain probed via the scanning probe could reveal valuable information on the local ionic species, concentration, and diffusivity. Thus, the technique could provide a window into local electrochemistry at the nanoscale. In ESM, the ionic oscillation is driven by the electrical potential [33], while in STIM, it is driven by the local thermal stress.

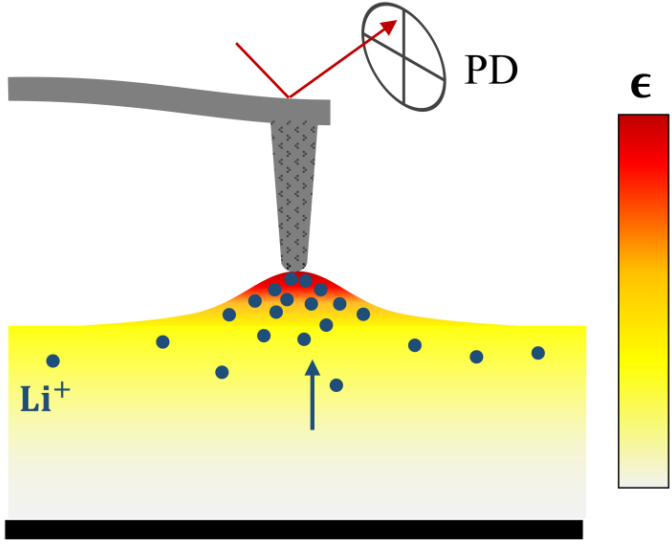

**Figure 1.** Schematic of Vegard strain detection in scanning probe microscopy. Higher ionic concentration induced by changes in electrochemical potential results in expanded molar volume that can be measured from the deflection of a scanning probe and the reflected laser beam, received in a photo detector (PD).

In order to fully appreciate the concept of STIM, it is necessary to briefly review the theory of stress-induced diffusion developed by Larché and Cahn in the 1970s [34, 35],

$$\frac{\partial c}{\partial t} = \nabla \cdot (D \nabla c) + \nabla \cdot \left(\frac{DFz}{RT} c \nabla \phi\right) - \nabla \cdot \left(\frac{D\Omega}{RT} c \nabla \sigma_h\right), \quad (1)$$

where $D$, $z$, and $\Omega$ are the diffusivity, charge, and partial molar volume of an ion, $F$ and $R$ are the Faraday and ideal gas constants, and $T$ and $t$ are the absolute temperature and time, respectively. Again, the three driving forces for ionic oscillation discussed earlier - gradients in ionic concentration $c$, electric potential $\phi$, and hydrostatic mechanical stress $\sigma_h$ - correspond to three terms on the right side of Equation (1). In order to impose an oscillating stress while simultaneously measuring the resulting local vibration through the scanning probe, we resort to a local temperature oscillation with angular frequency $\omega$,

$$\Delta T_{AC}[\omega] = \Delta T e^{i\omega t}, \quad (2)$$

which results in local oscillation of thermal strain $\Delta \boldsymbol{\varepsilon}^*$ and hydrostatic stress $\Delta \boldsymbol{\sigma_h}$,

$$\Delta \boldsymbol{\varepsilon}^*[\omega] = \alpha \Delta T e^{i\omega t} \mathbf{I}, \qquad \Delta \sigma_h[\omega] = \frac{1}{3} tr \mathbf{C}(\Delta \boldsymbol{\varepsilon}[\omega] - \Delta \boldsymbol{\varepsilon}^*[\omega]) = \Delta \sigma_{h0} e^{i\omega t}, \quad (3)$$

where $\alpha$ and $\mathbf{C}$ are the thermal expansion coefficient and elastic stiffness tensor of the material, $\boldsymbol{\varepsilon}$ is the total strain consisting of thermal strain $\boldsymbol{\varepsilon}^*$ and elastic strain, $tr$ denotes the trace of the matrix, $\sigma_{h0}$ is the amplitude of hydrostatic stress oscillation, and $\mathbf{I}$ is the second rank unit tensor. Note that thermal expansion results in a vibration that is the first harmonic to the temperature oscillation, as schematically shown in Figure 2. Measuring this first harmonic displacement response $u_1[\omega]$ reveals the local thermomechanical properties of the material.

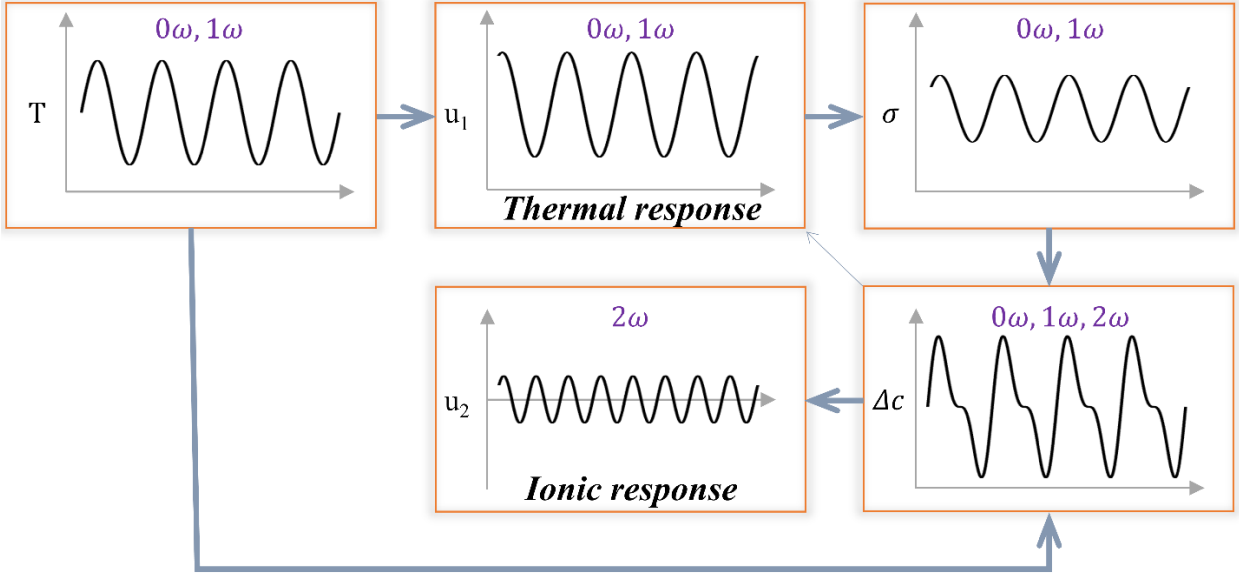

**Figure 2.** Graphical representations of oscillating temperature, stress, ionic concentration, and the first and second harmonic components of scanning thermo-ionic microscopy displacements. The oscillation of temperature causes thermal expansion that induces a fluctuating thermal stress. The simultaneous temperature and stress oscillations lead in an ionic concentration oscillation and subsequently displacement oscillation at a few distinctive frequencies.

However, there are further consequences and implications of this oscillating thermal stress, which drives oscillation in ionic concentration. By Taylor expanding $T$ into the series around the baseline temperature $T_0$ and upon substitution into Equation (1), we obtain the first and the second harmonic components of ionic oscillation:

$$\Delta c[\omega] = -\nabla \cdot \left(\frac{D\Omega c_0}{RT_0} \nabla \sigma_{h0}\right) e^{i\omega t}, \qquad \Delta c[2\omega] = \nabla \cdot \left(\frac{D\Omega c_0}{RT_0^2} \Delta T \nabla \sigma_{h0}\right) e^{i2\omega t}, \qquad (4),$$

which in turn induce the first and second harmonic Vegard strains and the corresponding displacements shown in Figure 2. Thus the first harmonic STIM response consists of contributions from both thermal expansion and Vegard strain, which is usually dominated by thermal expansion and thereafter referred to as *thermal response*, while the second harmonic STIM response is purely caused by Vegard strain associated with ionic oscillation and is thereafter referred to as *ionic response*. By measuring displacements associated with thermal and ionic responses induced by the fluctuating temperature and thermal stress at respective harmonics, information on both thermomechanical and ionic properties of materials can be obtained. This is the principle of STIM, as we have reported in Ref. [25] with preliminary implementation.

**Implementation.** Our initial implementation of STIM is to impose a temperature oscillation through a scanning thermal probe (AN2-300, Anasys Instruments), as shown in Figure 3a, on an Asylum Research MFP-3D AFM. The two-leg thermal probe has a micro-fabricated solid state resistive heater at the end of the cantilever, enabling local heating and thus temperature oscillation when an AC current is passed. In this implementation the power dissipation, and thus the temperature oscillation, are the second harmonic with respect to the AC current because of the quadratic relationship between the power and current. As a result, with reference to the input current, we would get a thermal response at the second harmonic and an ionic response at the fourth harmonic instead, which reflect the thermomechanical and electrochemical properties of the material, respectively.

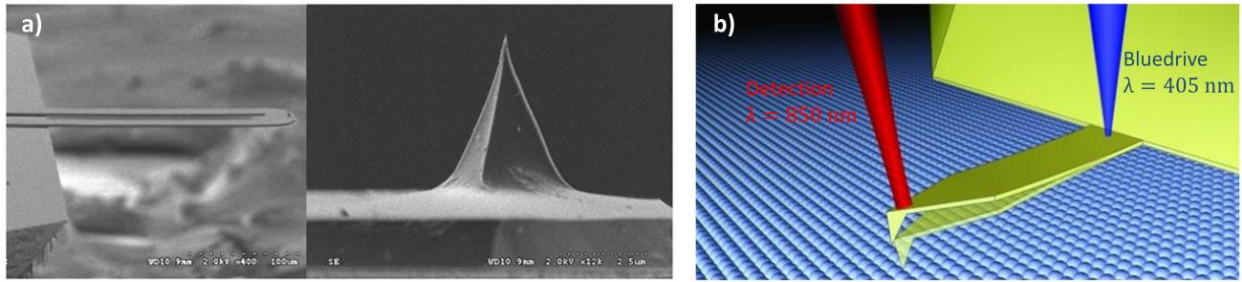

**Figure 3.** (a) Photos of two-leg scanning thermal probe and (b) blueDrive™ photothermal excitation module. Images used with permission of ANASYS Instruments and Asylum Research.

Alternatively, local heating and temperature fluctuation can also be realized through a photo-thermal approach, utilizing a 405 nm laser with modulated intensity aligned at the base of a gold coated cantilever (Multi75GD-G, Budget Sensors), implemented on an Asylum Research Cypher ES AFM equipped with blueDrive™ photo-thermal excitation module [36], as shown in Figure 3b. Under photo-thermal excitation, the laser power and thus the local temperature is modulated directly, and as such the first harmonic thermal response and second harmonic ionic responses will be obtained, as originally developed in Equations (4). We have implemented both approaches in our labs at the University of Washington (UW).

Detection of the respective harmonic response of the cantilever deflection, usually very small in magnitude, is accomplished by a lock-in amplifier around the cantilever-sample contact resonance frequency, which enhances the signal to noise ratio by orders of magnitude. To avoid topography cross-talk during STIM scanning, a dual amplitude resonance tracking (DART) technique [37], initially termed as dual frequency resonant tracking (DFRT), is used to track the contact resonance, as shown in Figure 4 implemented with the thermal probe. Four lock-ins are thus necessary to obtain the thermal (off-resonance) and ionic response (resonance-enhanced) mappings simultaneously via photo-thermal or Joule heating excitation, which is implemented using a Zurich Instrument HF2LI lock-in amplifier and proportional–integral–derivative (PID) controller in combination with an Asylum Research MFP-3D Bio AFM in the Shenzhen Key Laboratory of Nanobiomechanics, Shenzhen Institute of Advanced Technology, Chinese Academy of Science. Furthermore, the responses at two frequencies across the resonance allow us to solve for the quality factor and the contact resonance frequency of the system based on damped driven simple harmonic oscillator model [38, 39], enabling more accurate quantitative analysis. Lacking such a system at UW, resonance tracking is not possible for STIM, and single frequency scanning is used, which often causes cross-talk with topography due to inherent variation of material properties at the nanoscale. In such a case, point-wise data of respective harmonic responses will be more reliable, and spectroscopic studies using a combination of AC and DC voltage are also possible, as we show later.

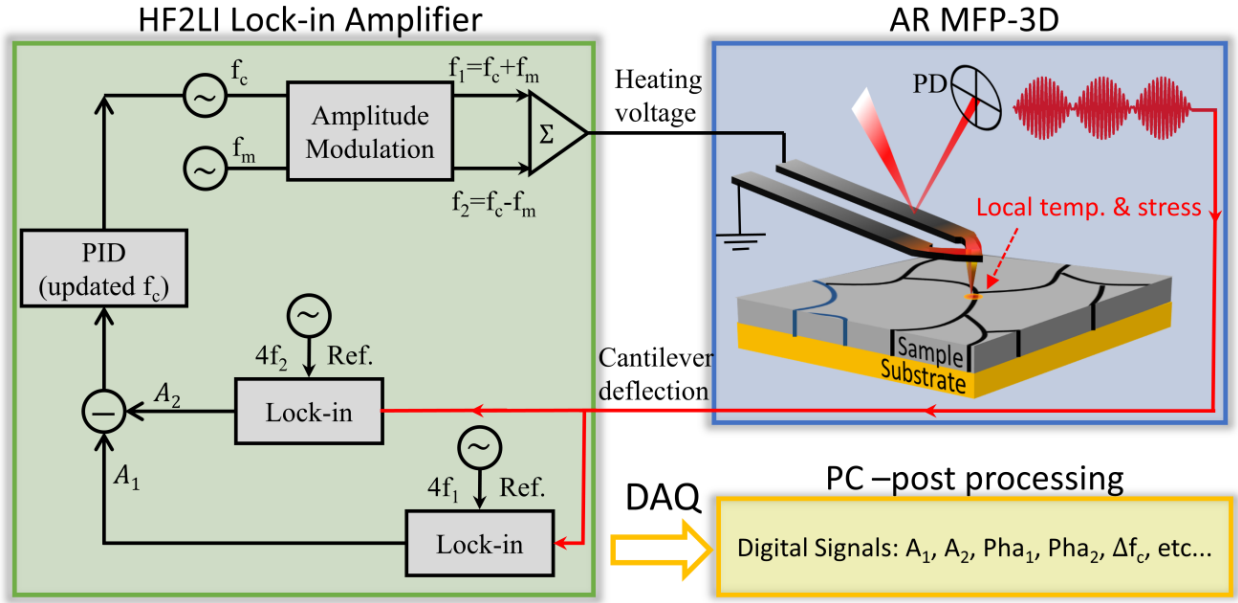

**Figure 4.** Schematic implementation of the dual amplitude resonance tracking (DART) scanning thermo-ionic microscopy (STIM). The lock-in amplifier synthesizes a dual AC signal (at frequencies $f_1$ and $f_2$) around the contact resonance of the probe, and the output of the lock-in is used for bimodal excitation of thermal probe. The AFM deflection signal is routed to the lock-in input and the 4$^{th}$ harmonic deflection responses at both frequencies are detected (amplitude and phase). The internal PID is used to regulate the difference between the detected amplitudes and adjust the frequency of the drive (carrier frequency $f_c$). The lock-in amplifier streams the detected information to a PC synchronous to AFM in real-time.

**Results**

   **Point-wise studies.** We first demonstrate the feasibility of the STIM concept by examining the point-wise thermal and ionic responses of two types of samples at respective harmonics. One is nanocrystalline Sm-doped ceria that is a good ionic conductor [40], and the other is polymeric polytetrafluoroethylene (PTFE) that serves as a control sample (no ionic conductivity). In the experiment, we sweep the driving frequency across probe-sample contact resonance, and measure the corresponding harmonic responses of probe deflection. As seen in Figure 5a, both ceria and PTFE exhibit substantial thermal response measured at the second harmonic under the resistive heating, as expected in any material regardless of its ionic characteristics. The higher contact resonance frequency of ceria of 123.4 kHz indicates its higher elastic modulus compared to PTFE with a resonance frequency of 118.7 kHz. Furthermore, after fitting the experimental data with the damped driven simple harmonic oscillator model [38, 39], the quality factors for ceria and PTFE are obtained as 57.7 and 46.4, which indicate higher viscous energy dissipation in PTFE. The corresponding intrinsic thermal response amplitudes (after correcting the peak value using the quality factors) are determined to be 38.2 pm and 44.3 pm, respectively, suggesting that the thermal expansion of PTFE is higher than that of ceria, as expected [41, 42]. Thus the local thermomechanical properties of solid materials can indeed be obtained from the STIM thermal response.

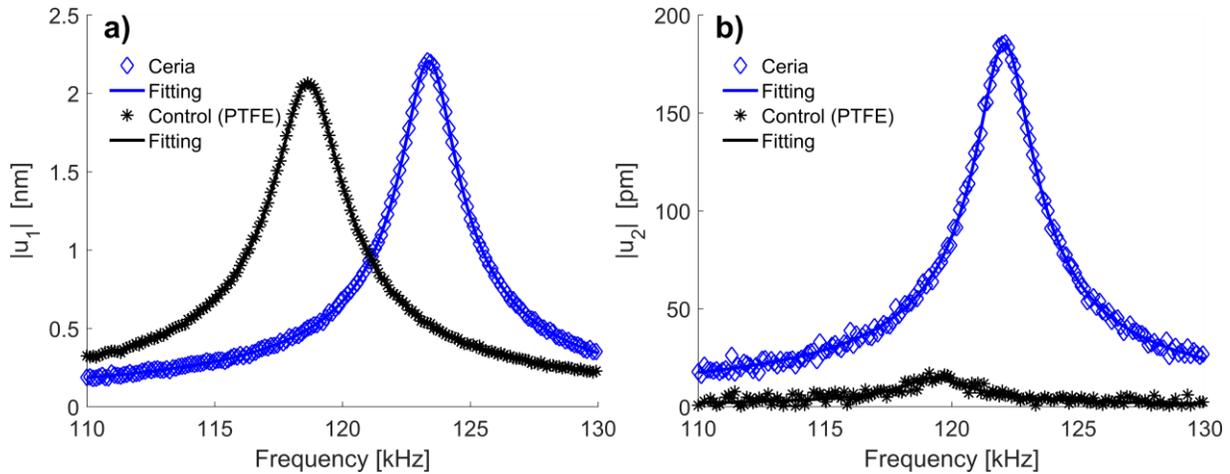

**Figure 5.** Comparison of point-wise (a) thermal and (b) ionic responses of ceria and PTFE samples measured using resistive heating on MFP-3D AFM, demonstrating the feasibility of STIM. While sweeping the drive signal frequency, the amplitude of (a) $2^{nd}$ and (b) $4^{th}$ harmonic response of the deflection signal is detected. The $2^{nd}$ and $4^{th}$ harmonic deflection signals are enhanced substantially where the temperature/ionic concentration oscillations (if any) are near the probe contact resonance.

Further insights can be learned from the ionic response measured at the fourth harmonic under similar conditions, as shown in Figure 5b. First of all, substantial higher ionic response is observed in ceria, while that of PTFE is almost negligible. This is more evident after we analyze the data using a damped driven simple harmonic model that yields a quality factor of 49.52 for PTFE and corresponding intrinsic ionic response of 0.31 pm, consistent with its non-ionic nature. The quality factor and the intrinsic response of ceria, on the other hand, are 55.95 and 3.31 pm. Since ionic response is proportional to ionic diffusivity according to Equation (4), the good ionic response of ceria seen in Figure 5b thus indicates its good ionic conductivity. Moreover, the quality factors and the resonance frequencies (119.3 and 122.1 kHz for PTFE and ceria) obtained from thermal and ionic responses are in a good agreement, confirming the reliability of the measurements. While the ionic response of ceria is an order of magnitude smaller compared to its thermal response, the signal is sufficiently strong and clean for sensitive detection. These sets of data thus demonstrate the feasibility of STIM.

We also can learn valuable dynamic information from the STIM ionic response, by imposing a low-frequency modulation bias, in the range of 0.1 Hz to 20 Hz, on top of high-frequency excitation bias, in the order of 30 kHz, as schematically shown in Figure 6a. The modulation bias is to manipulate the local ionic concentration away from equilibrium by changing the baseline temperature, while excitation bias is to stimulate ionic oscillation for the measurement. Since their frequencies differ by more than three orders of magnitude, the low-frequency modulation bias can be viewed as DC, as far as the instantaneous high-frequency excitation bias is concerned. The ionic responses of ceria measured in term of $4^{th}$ harmonic probe deflection induced by driving heating voltage, after correction by using a damped driven simple harmonic oscillator model, are plotted in Figure 6b as a function of DC voltage at different modulation frequencies, and different hysteresis loops are observed. It is observed that with increased DC voltage and thus higher local temperature, the ionic response drops due to reduced ionic concentration underneath the probe, driven away by the higher temperature and thermal stress. Furthermore, the loop initially opens up with an increased modulation in frequency and then closes, which can be understood as follows. For the low (0.2 Hz) and fast (20 Hz) modulation frequencies, where the ionic species are being driven much slower and faster than the typical diffusion time scale, not much hysteresis is observed. However, the intermediate frequency modulations (2 Hz), which has a time scale comparable to that of ionic diffusion, leads a phase lag and thus the loop opens up [43]. As such, from the loop opening under appropriate modulating frequency, dynamic information can be learned from the STIM spectroscopy studies.

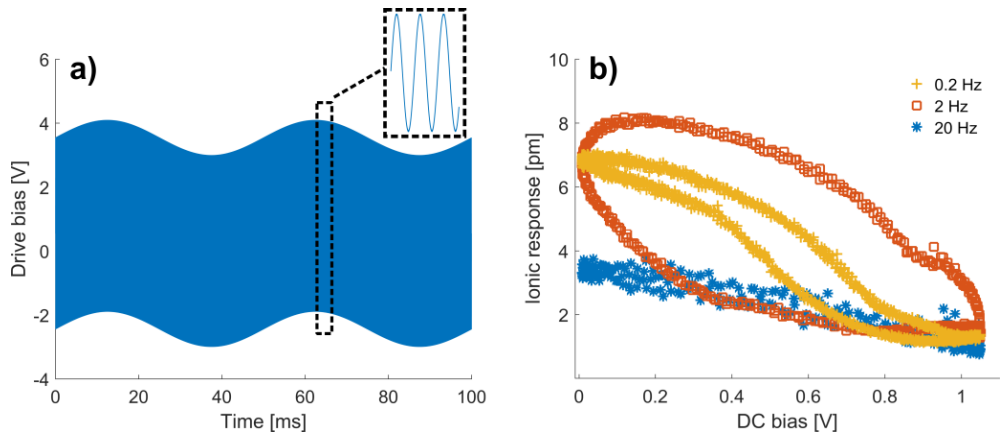

**Figure 6.** Point-wise temperature spectroscopy of ceria; (a) heating voltage applied to the thermal probe, consisting of superposition of high- and low-frequency voltage waveforms; (b) the intrinsic ionic response of ceria as a function low-frequency modulation bias under different modulating frequencies; these measurements were repeated 20 times and the average results are presented.

**Mapping.** The STIM mapping of thermal and ionic responses can reveal local variations of thermomechanical and electrochemical properties of materials, again using ceria as an example. Using DART STIM implemented on an MFP-3D Bio with HF2LI lock-in amplifier, thermal and ionic mappings of ceria have been acquired in a 900 nm × 900 nm region along with its contact mode topography. The top row of Figure 7 corresponds to thermal mapping while bottom row shows ionic maps. From topography mappings in Figures 7a and 7d, a pentagonal grain having boundaries with five neighboring grains is evident. The thermal response in Figure 7b shows grain-to-grain variation, as well as variation within a grain, possibly caused by the effect of topography variations on heat transfer between the thermal probe and sample. The ionic response in Figure 7e, on the other hand, exhibits a substantially higher amplitude at the grain boundaries, but within each grain the mappings are rather uniform. Such contrast is believed to be caused by accumulation of mobile electrons in the diffuse space charge regions near the surface and at grain boundaries, as recently imaged by ESM [26, 43, 44]. The resonance frequency mappings in Figure 7c and 7f, acquired from the second and fourth harmonic scanning are consistent with each other, though the second harmonic one has a better signal-to-noise ratio due to its much higher amplitude. Compared to what we originally reported [25], much improved STIM mappings are obtained, thanks to the improved capability for tracking the resonance frequency during scanning using DART.

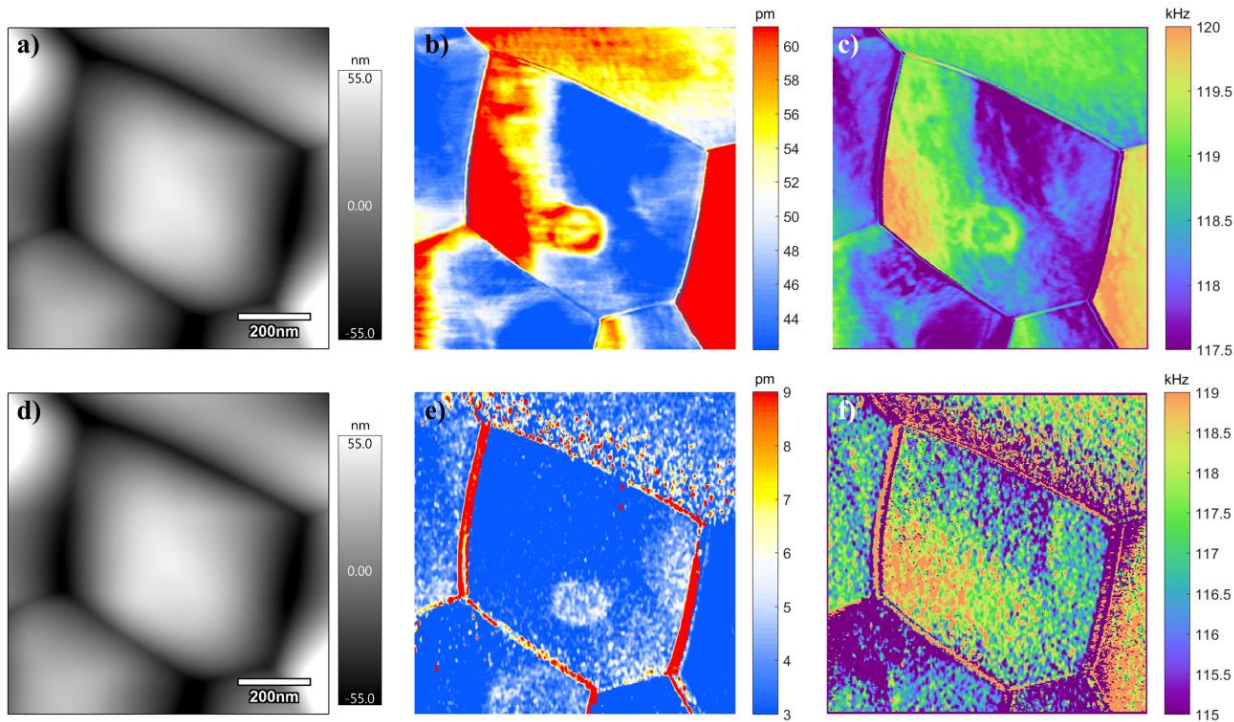

**Figure 7.** Thermomechanical (a-c) and electrochemical (d-f) responses of ceria acquired under dual amplitude resonance tracking (DART) scanning thermo-ionic microscopy (STIM) with resistive heating using a thermal probe, with 2$^{nd}$ harmonic in first row and 4$^{th}$ harmonic in second row; (a,d) topography; (b,e) thermal and ionic response amplitudes, and (c,f) resonance frequencies. The excitation frequencies were half and quarter of contact resonance frequency for thermal and ionic imaging, respectively.

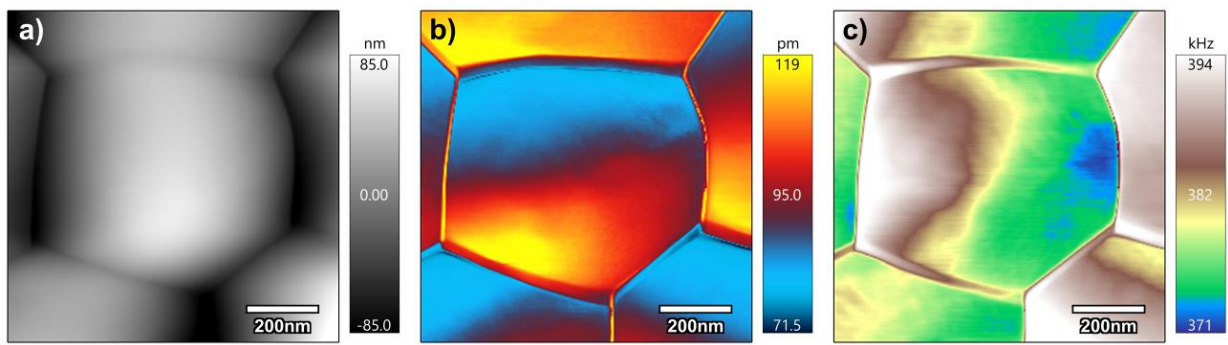

**Figure 8.** DART STIM of ceria acquired using a blueDrive™ module in Cypher ES AFM. (a) topography, (b) corrected thermal response amplitude, and (c) contact resonance frequency mappings.

The STIM mappings of ceria also were obtained using photo-thermal excitation through a contact mode scanning with a gold coated AFM probe (Multi75GD-G, Budget Sensors), as shown in Figure 8 around a pentagonal grain. Variation of thermal response in a grain shown in Figure 8b is similar to what we observed in Figure 7b, suggesting possible influence of topography on heat conduction, though the frequency variation in Figure 7c exhibits a different type of variation within the grain. The mechanism of such contrast is currently under investigation. The ionic image requires second harmonic resonance tracking which is currently unavailable in our Cypher ES system, but the mappings reveal contrasts of thermal expansion (Figure 8b) and mechanical stiffness (Figure 8c) across grains and grain boundaries.

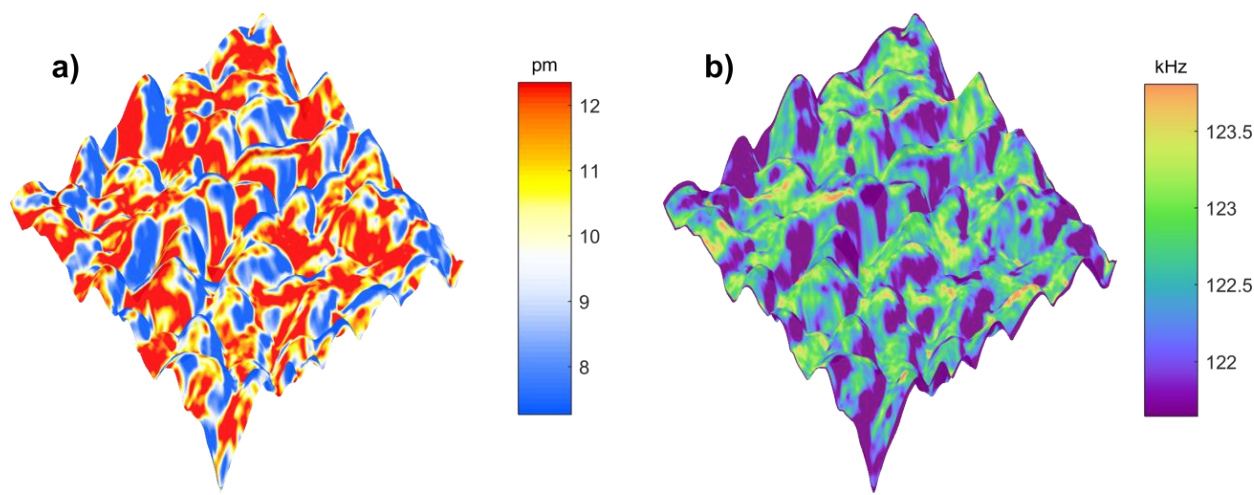

**Figure 9.** DART STIM mappings of perovskite $CH_3NH_3PbI_3$ acquired using a heated thermal probe on a MFP-3D Bio AFM. (a) the corrected ionic response amplitude and (b) the contact resonance frequency mappings, both overlaid on 3D topography in a $1\ \mu m \times\ 1\ \mu m$ area.

Finally, we show our preliminary results of STIM ionic mapping on $CH_3NH_3PbI_3$, an emerging perovskite material for solar cells [45, 46]. It has been reported that ionic motion in $CH_3NH_3PbI_3$ occurs under light illumination or electric field [47], which has significant implication to its photovoltaic performance. The STIM provides a powerful tool to study such phenomena locally at the nanoscale, as shown in Fig. 9 for a $CH_3NH_3PbI_3$ film on a hole-collecting FTO/PEDOTS:PSS substrate, in which rough topography is observed and large variations in ionic response are evident in Figure 9a. The resonant frequency shown in Figure 9b, on the other hand, is rather uniform, indicating homogeneous mechanical stiffness of the sample.

**Discussion**

Measuring ionic concentration and diffusivity with nanometer resolution is rather challenging using conventional electrochemical techniques. Scanning thermos-ionic microscopy (STIM), on the other hand, overcomes this difficulty by measuring Vegard strain arising from local ionic oscillation. Spatial resolution as small as 10 nm and sensitivity as high as picometer can be realized using a scanning probe, and the ionic response, measured in terms of probe deflection, correlates with local ionic concentration and diffusivity. It thus gives us a powerful tool to probe local electrochemistry at the nanoscale, and can be applied to study a wide range of electrochemical materials and systems wherein ionic activities are essential. Furthermore, thermal expansion and mechanical stiffness of the sample can also be obtained, yielding additional insight into the local electrochemical processes, for example the formation of solid electrolyte interface (SEI) in lithium ion batteries. The data analysis and interpretation of STIM, however, are not straightforward, and often requires assistance of modeling and simulation to translate the measured probe deflection into ionic concentration and diffusivity. In particular, spectroscopy type of dynamic studies are necessary to deconvolute local diffusivity and ionic concentration. Such capabilities are currently under development in our group, and will be reported later.

**Conclusion**

In this work, we demonstrate the capabilities of scanning thermo-ionic microscopy (STIM) for characterizing ionic and thermal properties of materials at nanoscale, based on detecting dynamic strain due to ionic motion induced by modulated temperature and stress gradients. The technique was demonstrated on active electrochemical materials including ceria and perovskite $CH_3NH_3PbI_3$ as well as a control sample of PTFE, implemented using both resistive heating and photo-thermal heating. The dynamics of ionic motion can be captured from point-wise spectroscopy studies, while the spatial inhomogeneity can be revealed by STIM mapping. Since it utilizes thermal stress-induced oscillation as its driving force, the responses are insensitive to the electromechanical, electrostatic, and capacitive effects, and they are immune to global current perturbation, making *in-operando* testing possible. In principle, STIM can provide a powerful tool for probing local electrochemical functionalities at the nanoscale.

**Acknowledgements**

This material is based in part upon work supported by National Key Research and Development Program of China (2016YFA0201001), National Natural Science Foundation of China (11627801, 11472236, 11772207), the Leading Talents Program of Guangdong Province (2016LJ06C372), and the State of Washington through the University of Washington Clean Energy Institute and via funding from the Washington Research Foundation.

**References**

1. Tarascon, J.M. and M. Armand, *Issues and challenges facing rechargeable lithium batteries.* Nature, 2001. **414**(6861): p. 359-67.
2. Dunn, B., H. Kamath, and J.M. Tarascon, *Electrical energy storage for the grid: a battery of choices.* Science, 2011. **334**(6058): p. 928-35.
3. Whittingham, M.S., *Lithium batteries and cathode materials.* Chem Rev, 2004. **104**(10): p. 4271-301.
4. Borup, R., et al., *Scientific aspects of polymer electrolyte fuel cell durability and degradation.* Chem Rev, 2007. **107**(10): p. 3904-51.
5. Adler, S.B., *Factors governing oxygen reduction in solid oxide fuel cell cathodes.* Chem Rev, 2004. **104**(10): p. 4791-843.
6. Kim, S., et al., *Cycling performance and efficiency of sulfonated poly(sulfone) membranes in vanadium redox flow batteries.* Electrochemistry Communications, 2010. **12**(11): p. 1650-1653.
7. Graves, C., et al., *Sustainable hydrocarbon fuels by recycling CO2 and H2O with renewable or nuclear energy.* Renewable & Sustainable Energy Reviews, 2011. **15**(1): p. 1-23.
8. Pinaud, B.A., et al., *Technical and economic feasibility of centralized facilities for solar hydrogen production via photocatalysis and photoelectrochemistry.* Energy & Environmental Science, 2013. **6**(7): p. 1983-2002.
9. Arico, A.S., et al., *Nanostructured materials for advanced energy conversion and storage devices.* Nat Mater, 2005. **4**(5): p. 366-77.
10. Groves, C., O.G. Reid, and D.S. Ginger, *Heterogeneity in polymer solar cells: local morphology and performance in organic photovoltaics studied with scanning probe microscopy.* Acc Chem Res, 2010. **43**(5): p. 612-20.
11. Kalinin, S.V. and N. Balke, *Local electrochemical functionality in energy storage materials and devices by scanning probe microscopies: status and perspectives.* Adv Mater, 2010. **22**(35): p. E193-209.
12. Huang, Z., et al., *PeakForce Scanning Electrochemical Microscopy with Nanoelectrode Probes.* Microscopy Today, 2016. **24**(6): p. 18-25.
13. Bard, A.J. and M.V. Mirkin, *Scanning electrochemical microscopy*. 2012: CRC Press.
14. Barker, A.L., et al., *Scanning electrochemical microscopy: beyond the solid/liquid interface.* Analytica Chimica Acta, 1999. **385**(1-3): p. 223-240.
15. Kalinin, S.V. and A. Gruverman, *Scanning probe microscopy: electrical and electromechanical phenomena at the nanoscale*. Vol. 1. 2007: Springer Science & Business Media.
16. Denton, A.R. and N.W. Ashcroft, *Vegard's law.* Phys Rev A, 1991. **43**(6): p. 3161-3164.
17. Vegard, L., *Die konstitution der mischkristalle und die raumfüllung der atome.* Zeitschrift für Physik, 1921. **5**(1): p. 17-26.

18. Tian, Y., A. Timmons, and J.R. Dahn, *In Situ AFM Measurements of the Expansion of Nanostructured Sn-Co-C Films Reacting with Lithium.* Journal of the Electrochemical Society, 2009. **156**(3): p. A187-A191.
19. Balke, N., et al., *Real space mapping of Li-ion transport in amorphous Si anodes with nanometer resolution.* Nano Lett, 2010. **10**(9): p. 3420-5.
20. Chen, Q.N., et al., *Delineating local electromigration for nanoscale probing of lithium ion intercalation and extraction by electrochemical strain microscopy.* Applied Physics Letters, 2012. **101**(6): p. 063901.
21. Zhu, J., et al., *In situ study of topography, phase and volume changes of titanium dioxide anode in all-solid-state thin film lithium-ion battery by biased scanning probe microscopy.* Journal of Power Sources, 2012. **197**: p. 224-230.
22. Eshghinejad, A., et al., *Resolving Local Electrochemistry at the Nanoscale via Electrochemical Strain Microscopy: Modeling and Experiments.* arXiv preprint arXiv:1704.01158, 2017.
23. Chen, Q.N., et al., *Mechanisms of electromechanical coupling in strain based scanning probe microscopy.* Applied Physics Letters, 2014. **104**(24): p. 242907.
24. Proksch, R., *Electrochemical strain microscopy of silica glasses.* Journal of Applied Physics, 2014. **116**(6): p. 066804.
25. Eshghinejad, A., et al., *Scanning thermo-ionic microscopy for probing local electrochemistry at the nanoscale.* Journal of Applied Physics, 2016. **119**(20): p. 205110.
26. Adler, S.B., et al., *Imaging Space Charge Regions in Sm-Doped Ceria Using Strain-Based Scanning Probe Techniques.* ECS Transactions, 2017. **78**(1): p. 335-342.
27. Kalinin, S.V. and D.A. Bonnell, *Imaging mechanism of piezoresponse force microscopy of ferroelectric surfaces.* Physical Review B, 2002. **65**(12): p. 125408.
28. Kolosov, O., et al., *Nanoscale visualization and control of ferroelectric domains by atomic force microscopy.* Phys Rev Lett, 1995. **74**(21): p. 4309-4312.
29. Keiji, T., et al., *Strain Imaging of Lead-Zirconate-Titanate Thin-Film by Tunneling Acoustic Microscopy.* Japanese Journal of Applied Physics Part 1-Regular Papers Short Notes & Review Papers, 1994. **33**(5b): p. 3193-3196.
30. Jiang, P., et al., *Electromechanical Coupling of Murine Lung Tissues Probed by Piezoresponse Force Microscopy.* Acs Biomaterials Science & Engineering, 2017. **3**(8): p. 1827-1835.
31. Chen, Q.N., et al., *High sensitivity piezomagnetic force microscopy for quantitative probing of magnetic materials at the nanoscale.* Nanoscale, 2013. **5**(13): p. 5747-51.
32. Eshghinejad, A., et al., *Piezoelectric and piezomagnetic force microscopies of multiferroic BiFeO3-LiMn2O4 heterostructures.* Journal of Applied Physics, 2014. **116**(6): p. 066805.
33. Morozovska, A.N., et al., *Local probing of ionic diffusion by electrochemical strain microscopy: Spatial resolution and signal formation mechanisms.* Journal of Applied Physics, 2010. **108**(5): p. 053712.
34. Larché, F. and J.W. Cahn, *A linear theory of thermochemical equilibrium of solids under stress.* Acta Metallurgica, 1973. **21**(8): p. 1051-1063.
35. Eshghinejad, A. and J.Y. Li, *The coupled lithium ion diffusion and stress in battery electrodes.* Mechanics of Materials, 2015. **91**: p. 343-350.
36. Labuda, A., et al., *Photothermal excitation for improved cantilever drive performance in tapping mode atomic force microscopy.* Microscopy and Analysis28, 2014. **3**: p. S21-S25.
37. Rodriguez, B.J., et al., *Dual-frequency resonance-tracking atomic force microscopy.* Nanotechnology, 2007. **18**(47): p. 475504.
38. Liu, Y.Y., et al., *Controlling magnetoelectric coupling by nanoscale phase transformation in strain engineered bismuth ferrite.* Nanoscale, 2012. **4**(10): p. 3175-83.
39. French, A.P., *Vibrations and waves*. 2001, AAPT.

40. Tuller, H.L., *Ionic conduction in nanocrystalline materials.* Solid State Ionics, 2000. **131**(1-2): p. 143-157.
41. Hayashi, H., et al., *Thermal expansion of Gd-doped ceria and reduced ceria.* Solid State Ionics, 2000. **132**(3-4): p. 227-233.
42. Kirby, R.K., *Thermal expansion of polytetrafluoroethylene (Teflon) from− 190° to+ 300° C.* Journal of Research of the National Bureau of Standards, 1956. **57**(2): p. 91-94.
43. Chen, Q.N., S.B. Adler, and J.Y. Li, *Imaging space charge regions in Sm-doped ceria using electrochemical strain microscopy.* Applied Physics Letters, 2014. **105**(20): p. 201602.
44. Li, J., et al., *Strain-based scanning probe microscopies for functional materials, biological structures, and electrochemical systems.* Journal of Materiomics, 2015. **1**(1): p. 3-21.
45. Fang, R., et al., *The rising star in photovoltaics-perovskite solar cells: The past, present and future.* Science China-Technological Sciences, 2016. **59**(7): p. 989-1006.
46. Wang, P., et al., *Photo-induced ferroelectric switching in perovskite CH 3 NH 3 PbI 3 films.* Nanoscale, 2017.
47. Xu, X. and M. Wang, *Photocurrent hysteresis related to ion motion in metal-organic perovskites.* Science China Chemistry, 2016: p. 1-9.